\documentclass[review]{elsarticle}

\usepackage{amsmath}
\usepackage{ragged2e}
\journal{Journal of \LaTeX\ Templates}

\begin{document}

\begin{frontmatter}

\title{Geometric Moment Invariants to Motion Blur}


\author[mymainaddress,mysecondaryaddress]{Hongxiang Hao.\corref{mycorrespondingauthor}}
\cortext[mycorrespondingauthor]{Corresponding author}
\ead{haohongxiang18s@ict.ac.cn}

\author[mymainaddress,mysecondaryaddress]{Hanlin Mo.}
\author[mymainaddress,mysecondaryaddress]{Hua Li.}

\address[mymainaddress]{Key Laboratory of Intelligent Information Processing, Institute of \\Computing Technology, Chinese Academy of Sciences, Beijing, China}
\address[mysecondaryaddress]{University of Chinese Academy of Sciences, Beijing, China}

\begin{abstract}
In this paper, we focus on removing interference of motion blur by the derivation of motion blur invariants.Unlike earlier work, we don’t restore any blurred image. Based on geometric moment and mathematical model of motion blur, we prove that geometric moments of blurred image and original image are linearly related. Depending on this property, we can analyse whether an existing moment-based feature is invariant to motion blur. Surprisingly, we find some geometric moment invariants are invariants to not only spatial transform but also motion blur. Meanwhile, we test invariance and robustness of these invariants using synthetic and real blur image datasets. And the results show these invariants outperform some widely used blur moment invariants and non-moment image features in image retrieval, classification and template matching.
\end{abstract}

\begin{keyword}
\texttt invariance \sep geometric moment invariant \sep rotational motion blur \sep linear motion blur \sep  feature extraction \sep image retrieval
\end{keyword}

\end{frontmatter}

%%\linenumbers
\section{Introduction}

Image analysis and recognition is one of the most fundamental goals in many fields including computer vision. Researchers always expect high-quality images, but many factors can cause degradation and interference during the process of imaging. Among these factors, blur is the most common, including atmospheric interference, out of focus, relative motion between camera and scene, and other ways of degradation. Actually, blur is an irreversible process with the abnormal diffusion and superposition of image pixels. In general, image degradation can be described by the following mathematical model\cite{ref1}:
\begin{equation}
g(x',y')=f*h(x,y)+n(x,y)
\end{equation}
where h stands for the point-spread function (PSF) of the imaging system.

Motion blur is a common degradation. It can be modeled as the integration over time of image density function. According to form of relative motion between the object and the imaging system during exposure time, motion blur can be divided into 3 categories, such as linear motion blur, rotational motion blur and radial motion blur, which is more complex.

To remove interference of motion blur , most existing work focuses on the realization of "reversibility" and try to restore blurred image, named as “DEBLUR”. Although, many researchers have obtained excellent results and widely used them into many fields, they still faces many challenges. For example, complicated algorithm, heavy computation,probability of introducing new noises and so on.

In this paper, we adapt another approach of removing influences caused by motion blur, as known as extracting features that are invariant to motion blur. We focus on achieving “deblur-free” and conducting an ideal system of recognition. So our work mainly consists of three parts : First, we analyze the mathematical model of motion blur, combine it with the concept of geometric moments ; Then, we derive that geometric moments of blurred image and original image are linearly relative and screen out several moment invariants depending on the derivation ; Finally, we test invariance and robustness of these invariants using synthetic and real blur image datasets.

\clearpage
\section{Basic concepts and theories}
 In this section, we introduce some basic definitions and classical conclusions for our work.
\subsection{The mathematical model of motion blur}
Based on the model (1), we can analyze the PSF of motion blur and derive another model with integration over time instead of convolution. We find the model with integration can describe motion blur more directly and solve so much mathematical problems caused by the complex convolution.So in our work, the degradation model can be defined as\cite{ref2}:
\begin{equation}
g({x_{(g)}},{y_{(g)}}) = \frac{1}{T}\int_0^T {f({x_{(t)}},{y_{(t)}})} \,dt
\end{equation}
Where T stands for the exposure time, g stands for the blurred image and f stands for the image after corresponding spatial transform. In image density functions g ,$({x_{(g)}},{y_{(g)}})$ stands for point coordinates of blurred image; and $({x_{(t)}},{y_{(t)}})$ in f stands for point coordinates of image during spatial transform.

According to the model (2), we can regard linear motion blur as the diffusion and superposition of a series image translation transform.Supposing a and b stands for linear velocity of translation in x axis and in y axis direction. So the relations of point coordinates among original image, moving image and blurred image can be derived as:
\begin{equation}
\left\{ \begin{array}{l}
\begin{aligned}
{x_{(t)}} &= x + at\\
{y_{(t)}} &= y + bt
\end{aligned}
\end{array} \right.\\
\left\{ \begin{array}{l}
\begin{aligned}
{x_{(g)}} &= x + aT\\
{y_{(g)}} &= y + bT
\end{aligned}
\end{array} \right.\\
\left\{ \begin{array}{l}
\begin{aligned}
{x_{(g)}} &= {x_{(t)}}- at+aT\\
{y_{(g)}} &= {y_{(t)}}- bt+bT
\end{aligned}
\end{array} \right.
\end{equation}

When we focus on rotational motion blu, we can see it as the diffusion and superposition of a series image rotation transform.Supposing $\omega$ stands for angular velocity of rotation, $\alpha$ stands for deflection angle before and after rotation blur. So we can build the relations of point coordinates among original image, moving image and blurred image as: 
\begin{equation}
\begin{array}{l}
\begin{aligned}
\left[ {\begin{array}{*{20}{c}}
{{x_{(g)}}}\\
{{y_{(g)}}}
\end{array}} \right] &= \left[ {\begin{array}{*{20}{c}}
{\cos \alpha }&{ - \sin \alpha }\\
{\sin \alpha }&{\cos \alpha }
\end{array}} \right] \cdot \left[ {\begin{array}{*{20}{c}}
x\\
y
\end{array}} \right]\\
\left[ {\begin{array}{*{20}{c}}
{{x_{(t)}}}\\
{{y_{(t)}}}
\end{array}} \right] &= \left[ {\begin{array}{*{20}{c}}
{\cos {\phi _{(t)}}}&{ - \sin {\phi _{(t)}}}\\
{\sin {\phi _{(t)}}}&{\cos {\phi _{(t)}}}
\end{array}} \right] \cdot \left[ {\begin{array}{*{20}{c}}
x\\
y
\end{array}} \right]    \\
\left[ {\begin{array}{*{20}{c}}
{{x_{(t)}}}\\
{{y_{(t)}}}
\end{array}} \right]& = \left[ {\begin{array}{*{20}{c}}
{\cos {\theta _{(t)}}}&{ - \sin {\theta _{(t)}}}\\
{\sin {\theta _{(t)}}}&{\cos {\theta _{(t)}}}
\end{array}} \right] \cdot \left[ {\begin{array}{*{20}{c}}
{{x_{(g)}}}\\
{{y_{(g)}}}
\end{array}} \right]
\end{aligned}
\end{array}
\end{equation}
where ${\phi _{(t)}} = \omega t$ and ${\theta _{(t)}} = {\phi _{(t)}} - \alpha  = \omega t - \alpha $ .

\subsection{Geometric moments}
Geometric moments are regarded as the simplest image moments.For an image $f(x,y)$, its geometric moments of (p+q)th order can be defined as:
\begin{equation}
m_{pq}^{(f)} = \int {\int {{x^p}{y^q}f(x,y)}}\,dxdy
\end{equation}
where p and q are non-negative integers. Based on (5), we can define centroid of the image as:
\begin{equation}
{x_c} = \frac{{m_{10}^{(f)}}}{{m_{00}^{(f)}}}~~~~{y_c} = \frac{{m_{01}^{(f)}}}{{m_{00}^{(f)}}}
\end{equation}

When we move the center of the image to centroid and re-establish the coordinate system, the influence of translation transform can be easily removed. So we define geometric central moments of (p+q)th order as:
\begin{equation}
u_{pq}^{(f)} = \int {\int {{{(x - {x_c})}^p}{{(y - {y_c})}^q}f(x,y)}}\,dxdy
\end{equation}

It’s obvious that first central moments of any image equals zero according to definition (7).

\subsection{Geometric moment invariants}
In 1962, Hu first proposed the concept of image geometric moments\cite{ref3}. He employed the results of the theory of algebratic invariants and derived 7 geometric moment invariants to similarity transform (including translation,scale and rotation) as known as Hu moment invariants. 
\begin{justifying}
\begin{equation}
\begin{array}{l}
\begin{aligned}
{\varphi _1} &= {\eta _{20}} + {\eta _{02}};\\
{\varphi _2} &= {({\eta _{20}} - {\eta _{02}})^2} + 4{\eta _{11}}^2;\\
{\varphi _3} &= {({\eta _{30}} - 3{\eta _{12}})^2} + {(3{\eta _{21}} - {\eta _{03}})^2};\\
{\varphi _4} &= {({\eta _{30}} + {\eta _{12}})^2} + {({\eta _{21}} + {\eta _{03}})^2};\\
{\varphi _5} &= ({\eta _{30}} - 3{\eta _{12}})({\eta _{30}} + {\eta _{12}})\left[ {{{({\eta _{30}} + {\eta _{12}})}^2} - 3{{({\eta _{21}} + {\eta _{03}})}^2}} \right] + (3{\eta _{21}} - {\eta _{03}})\\&\cdot ({\eta _{21}} + {\eta _{03}})  \left[ {3{{({\eta _{30}} + {\eta _{12}})}^2} - {{({\eta _{21}} + {\eta _{03}})}^2}} \right];\\
{\varphi _6} &= ({\eta _{20}} - {\eta _{02}})\left[ {{{({\eta _{30}} + {\eta _{12}})}^2} - {{({\eta _{21}} + {\eta _{03}})}^2}} \right] + 4{\eta _{11}}({\eta _{30}} - {\eta _{03}})({\eta _{21}} + {\eta _{03}})\\&\cdot\left[ {3{{({\eta _{30}} + {\eta _{12}})}^2} - {{({\eta _{21}} + {\eta _{03}})}^2}} \right];\\
{\varphi _7} &= (3{\eta _{21}} - {\eta _{03}})({\eta _{30}} + {\eta _{12}})\left[ {{{({\eta _{30}} + {\eta _{12}})}^2} - 3{{({\eta _{21}} + {\eta _{03}})}^2}} \right] + ({\eta _{30}} - 3{\eta _{12}})\\&\cdot({\eta _{21}} + {\eta _{03}})\left[ {3{{({\eta _{30}} + {\eta _{12}})}^2} - {{({\eta _{21}} + {\eta _{03}})}^2}} \right];
\end{aligned}
\end{array}
\end{equation}
\end{justifying}

These 7 invariants have become much popular image descriptors and have found numerous applications, namely in shape analysis\cite{ref4}\cite{ref5}, object recognition\cite{ref6}\cite{ref7}, and speech analysis\cite{ref8}.

After being widely used for nearly forty years, Flusser found this system is not independent\cite{ref9}. They showed that 7th invariant can be expressed as the function of the others:
\begin{equation}
{(\varphi 7)^2} = \varphi 3 \cdot {(\varphi 4)^3} - {(\varphi 5)^2}
\end{equation}
which indicates that 7th invariant can be excluded from the set without any loss of discrimination power So we only discuss 1st - 6th invariants in this paper.

\clearpage
\section{Geometric moment invariants to linear motion blur}
Based on equation (3) and definition(2) (7), we can derive centroid of the blurred image as:
\begin{equation}
\begin{array}{c}
\begin{aligned}
{{\bar x}_{(g)}} &= \frac{{\mathop{{\int\!\!\!\!\!\int}\mkern-21mu \bigcirc} 
 {{x_{(g)}} \cdot g({x_{(g)}},{y_{(g)}})d{x_{(g)}}d{y_{(g)}}} }}{{\mathop{{\int\!\!\!\!\!\int}\mkern-21mu \bigcirc} 
 {g({x_{(g)}},{y_{(g)}})d{x_{(g)}}d{y_{(g)}}} }}\\
 &= \frac{{\mathop{{\int\!\!\!\!\!\int}\mkern-21mu \bigcirc} 
 {\left\{ {\left( {{x_{(t)}} - at + aT} \right) \cdot \left[ {\frac{1}{T}\int_0^T {f({x_{(t)}},{y_{(t)}})dt} } \right]} \right\}d{x_{(t)}}d{y_{(t)}}} }}{{\mathop{{\int\!\!\!\!\!\int}\mkern-21mu \bigcirc} 
 {f({x_{(t)}},{y_{(t)}})d{x_{(t)}}d{y_{(t)}}} }}\\
 &= \frac{{\frac{1}{T}\int_0^T {\left\{ {\left[ {\mathop{{\int\!\!\!\!\!\int}\mkern-21mu \bigcirc} 
 {\left[ {\left( {{x_{(t)}} - at + aT} \right) \cdot f({x_{(t)}},{y_{(t)}})} \right]d{x_{(t)}}d{y_{(t)}}} } \right]} \right\}dt} }}{{m_{00}^{(t)}}}\\
 &= \frac{{\frac{1}{T}\int_0^T {\left[ {m_{10}^{(t)} + \left( { - at + aT} \right) \cdot m_{00}^{(t)}} \right]} dt}}{{m_{00}^{(t)}}}\\
 &= {{\bar x}_{(t)}} + \frac{1}{2}aT\\\\
{{\bar y}_{(g)}} &= \frac{{\mathop{{\int\!\!\!\!\!\int}\mkern-21mu \bigcirc} 
 {{y_{(g)}} \cdot g({x_{(g)}},{y_{(g)}})d{x_{(g)}}d{y_{(g)}}} }}{{\mathop{{\int\!\!\!\!\!\int}\mkern-21mu \bigcirc} 
 {g({x_{(g)}},{y_{(g)}})d{x_{(g)}}d{y_{(g)}}} }}\\
 &= \frac{{\mathop{{\int\!\!\!\!\!\int}\mkern-21mu \bigcirc} 
 {\left\{ {\left( {{y_{(t)}} - bt + bT} \right) \cdot \left[ {\frac{1}{T}\int_0^T {f({x_{(t)}},{y_{(t)}})dt} } \right]} \right\}d{x_{(t)}}d{y_{(t)}}} }}{{\mathop{{\int\!\!\!\!\!\int}\mkern-21mu \bigcirc} 
 {f({x_{(t)}},{y_{(t)}})d{x_{(t)}}d{y_{(t)}}} }}\\
 &= \frac{{\frac{1}{T}\int_0^T {\left\{ {\left[ {\mathop{{\int\!\!\!\!\!\int}\mkern-21mu \bigcirc} 
 {\left[ {\left( {{y_{(t)}} - bt + bT} \right) \cdot f({x_{(t)}},{y_{(t)}})} \right]d{x_{(t)}}d{y_{(t)}}} } \right]} \right\}dt} }}{{m_{00}^{(t)}}}\\
 &= \frac{{\frac{1}{T}\int_0^T {\left[ {m_{01}^{(t)} + \left( { - bt + bT} \right) \cdot m_{00}^{(t)}} \right]} dt}}{{m_{00}^{(t)}}}\\
 &= {{\bar y}_{(t)}} + \frac{1}{2}bT
\end{aligned}
\end{array}
\end{equation}

\clearpage
Then we can define (p+q)th-order geometric central moments of blurred image as:
\begin{justifying}
\begin{equation}
\begin{array}{c}
\begin{aligned}
u_{pq}^{(g)} &= \mathop{{\int\!\!\!\!\!\int}\mkern-21mu \bigcirc} 
 {{{\left( {{x_{(g)}} - {{\bar x}_{(g)}}} \right)}^p}{{\left( {{y_{(g)}} - {{\bar y}_{(g)}}} \right)}^q}g({x_{(g)}},{y_{(g)}})d{x_{(g)}}d{y_{(g)}}} \\
 &= \mathop{{\int\!\!\!\!\!\int}\mkern-21mu \bigcirc} 
 {\left\{ {\frac{1}{T}\left[ {\int_0^T {f({x_{(t)}},{y_{(t)}})} dt} \right] \cdot {{\left( {{x_{(t)}} - {{\bar x}_{(t)}} - at + \frac{1}{2}aT} \right)}^p}} \right.} \\
&\left. { \cdot {{\left( {{y_{(t)}} - {{\bar y}_{(t)}} - bt + \frac{1}{2}bT} \right)}^q}} \right\}d{x_{(t)}}d{y_{(t)}}\\
 &= \frac{1}{T}\int_0^T {\left\{ {f({x_{(t)}},{y_{(t)}}) \cdot \mathop{{\int\!\!\!\!\!\int}\mkern-21mu \bigcirc} 
 {\left[ {{{\left( {{x_{(t)}} - {{\bar x}_{(t)}} - at + \frac{1}{2}aT} \right)}^p}} \right.} } \right.} \\
&\left. {\left. { \cdot {{\left( {{y_{(t)}} - {{\bar y}_{(t)}} - bt + \frac{1}{2}bT} \right)}^q}} \right]d{x_{(t)}}d{y_{(t)}}} \right\}dt\\
 &= \frac{1}{T}\int_0^T {\left\{ {\mathop{{\int\!\!\!\!\!\int}\mkern-21mu \bigcirc} 
 {\left[ {f({x_{(t)}},{y_{(t)}}) \cdot \sum\limits_{i = 0}^p {\left( {C_p^i \cdot {{\left( {{x_{(t)}} - {{\bar x}_{(t)}}} \right)}^i} \cdot {{\left( { - at + \frac{1}{2}aT} \right)}^{p - i}}} \right)} } \right.} } \right.} \\
&\left. {\left. { \cdot \sum\limits_{j = 0}^q {\left( {C_q^j \cdot {{\left( {{y_{(t)}} - {{\bar y}_{(t)}}} \right)}^j} \cdot {{\left( { - bt + \frac{1}{2}bT} \right)}^{q - j}}} \right)} } \right]d{x_{(t)}}d{y_{(t)}}} \right\}dt\\
 &= \frac{1}{T}\int_0^T {\left[ {\sum\limits_{i = j = 0}^{i \le p,j \le q} {\left( {u_{ij}^{(f)} \cdot C_p^i \cdot C_q^j \cdot {a^{p - i}} \cdot {b^{q - j}} \cdot {{(\frac{T}{2} - t)}^{p + q - i - j}}} \right)} } \right]} dt\\
 &= \sum\limits_{i = j = 0}^{i \le p,j \le q} {\left( {u_{ij}^{(f)} \cdot C_p^i \cdot C_q^j \cdot {a^{p - i}} \cdot {b^{q - j}} \cdot \frac{1}{T}\int_0^T {\left[ {{{(\frac{T}{2} - t)}^{p + q - i - j}}} \right]} dt} \right)} \\
 &= \sum\limits_{i = j = 0}^{i \le p,j \le q} {\left( {u_{ij}^{(f)} \cdot C_p^i \cdot C_q^j \cdot {a^{p - i}} \cdot {b^{q - j}} \cdot {{\left( {\frac{T}{2}} \right)}^{p + q - i - j}} \cdot \frac{{1 + {{\left( { - 1} \right)}^{p + q - i - j}}}}{{2 \cdot (p + q - i - j + 1)}}} \right)} 
\end{aligned}
\end{array}
\end{equation}
\end{justifying}

According the simplified result in the following definition (12), we can learn that geometric central moments of linear-motion-blurred image and original image are linearly relative.
\begin{equation}
\begin{array}{c}
\begin{aligned}
u_{pq}^{(g)} &= \sum\limits_{i = j = 0}^{i \le p,j \le q} {\left( {u_{ij}^{(f)} \cdot H(p,q,i,j)} \right)} \\
H(p,q,i,j) &= C_p^i \cdot C_q^j \cdot {a^{p - i}} \cdot {b^{q - j}} \cdot S(p,q,i,j)\\
S(p,q,i,j) &= \left\{ \begin{array}{l}
0~~~~~when~(p+q-i-j)~is~odd  \\
\frac{{{{\left( {{T \mathord{\left/
 {\vphantom {T 2}} \right.
 \kern-\nulldelimiterspace} 2}} \right)}^{p + q - i - j}}}}{{p + q - i - j + 1}}~~~~~when~(p+q-i-j)~is~even  
\end{array} \right.
\end{aligned}
\end{array}
\end{equation}

Based on the definition, we can obtain different moments of any order.In this paper, we give central moments from 1st to 4th order:
\begin{equation}
\begin{array}{c}
\begin{aligned}
1st:u_{10}^{(g)} &= 0\\u_{01}^{(g)} &= 0\\
2nd:u_{20}^{(g)} &= u_{20}^{(f)} + u_{00}^{(f)} \cdot \frac{1}{3} \cdot {\left( {\frac{{aT}}{2}} \right)^2}\\
u_{11}^{(g)} &= u_{11}^{(f)} + u_{00}^{(f)} \cdot \frac{1}{3} \cdot \frac{{aT}}{2} \cdot \frac{{bT}}{2}\\
u_{02}^{(g)} &= u_{02}^{(f)} + u_{00}^{(f)} \cdot \frac{1}{3} \cdot {\left( {\frac{{bT}}{2}} \right)^2}\\
3rd:u_{30}^{(g)} &= u_{30}^{(f)}\\u_{21}^{(g)} &= u_{21}^{(f)}\\u_{12}^{(g)} &= u_{12}^{(f)}\\u_{03}^{(g)} &= u_{03}^{(f)}\\
4th:u_{40}^{(g)} &= u_{40}^{(f)} + u_{20}^{(f)} \cdot 2 \cdot {\left( {\frac{{aT}}{2}} \right)^2} + u_{00}^{(f)} \cdot \frac{1}{5} \cdot {\left( {\frac{{aT}}{2}} \right)^4}\\
u_{31}^{(g)} &= u_{31}^{(f)} + u_{20}^{(f)} \cdot \frac{{aT}}{2} \cdot \frac{{bT}}{2} + u_{11}^{(f)} \cdot {\left( {\frac{{aT}}{2}} \right)^2} + u_{00}^{(f)} \cdot \frac{1}{5} \cdot {\left( {\frac{{aT}}{2}} \right)^3} \cdot \frac{{bT}}{2}\\
u_{22}^{(g)} &= u_{22}^{(f)} + u_{20}^{(f)} \cdot \frac{1}{3} \cdot {\left( {\frac{{bT}}{2}} \right)^2} + u_{02}^{(f)} \cdot \frac{1}{3} \cdot {\left( {\frac{{aT}}{2}} \right)^2} + u_{11}^{(f)} \cdot \frac{4}{3} \cdot \frac{{aT}}{2} \cdot \frac{{bT}}{2} \\&+ u_{00}^{(f)} \cdot \frac{1}{5} \cdot {\left( {\frac{{aT}}{2}} \right)^2} \cdot {\left( {\frac{{bT}}{2}} \right)^2}\\
u_{13}^{(g)} &= u_{13}^{(f)} + u_{02}^{(f)} \cdot \frac{{aT}}{2} \cdot \frac{{bT}}{2} + u_{11}^{(f)} \cdot {\left( {\frac{{bT}}{2}} \right)^2} + u_{00}^{(f)} \cdot \frac{1}{5} \cdot \frac{{aT}}{2} \cdot {\left( {\frac{{bT}}{2}} \right)^3}\\
u_{04}^{(g)} &= u_{04}^{(f)} + u_{02}^{(f)} \cdot 2 \cdot {\left( {\frac{{bT}}{2}} \right)^2} + u_{00}^{(f)} \cdot \frac{1}{5} \cdot {\left( {\frac{{bT}}{2}} \right)^4}
\end{aligned}
\end{array}
\end{equation}

According to the results shown in equation (13), we learn that four 3rd-order geometric central moments are natural invariants to linear motion blur. So any invariant that consists of only 3rd-order geometric central moments also has invariance to linear motion blur.

\section{Geometric moment invariants to rotational motion blur}
\subsection{Geometric moments of blurred image}
Based on equation (4) and definition(2)(7), we can define (p+q)th-order geometric central moments of blurred image as:
\begin{justifying}
\begin{equation}
\begin{array}{c}
\begin{aligned}
m_{pq}^{(g)} &= \mathop{{\int\!\!\!\!\!\int}\mkern-21mu \bigcirc} 
 {\left[ {\int {x_{(g)}^py_{(g)}^qg} ({x_{(g)}},{y_{(g)}})} \right]} d{x_{(g)}}d{y_{(g)}}\\
 &= \mathop{{\int\!\!\!\!\!\int}\mkern-21mu \bigcirc} 
 {\left\{ {x_{(g)}^py_{(g)}^q \cdot \left[ {\frac{1}{T}\int_0^T {f({x_{(t)}},{y_{(t)}})} dt} \right]} \right\}d{x_{(g)}}d{y_{(g)}}} \\
 &= \frac{1}{T}\int_0^T {\left\{ {\mathop{{\int\!\!\!\!\!\int}\mkern-21mu \bigcirc} 
 {\left[ {x_{(g)}^py_{(g)}^qf({x_{(t)}},{y_{(t)}})} \right]d{x_{(g)}}d{y_{(g)}}} } \right\}} dt\\
 &= \frac{1}{T}\int_0^T {\left\{ {\mathop{{\int\!\!\!\!\!\int}\mkern-21mu \bigcirc} 
 {\left[ {f({x_{(t)}},{y_{(t)}}) \cdot {{({x_{(t)}}\cos {\theta _{(t)}} + {y_{(t)}}\sin {\theta _{(t)}})}^p}} \right.} } \right.} \\
 &\cdot \left. {\left. {{{( - {x_{(t)}}\sin {\theta _{(t)}} + {y_{(t)}}\cos {\theta _{(t)}})}^q}} \right]d{x_{(t)}}d{y_{(t)}}} \right\}dt\\
 &= \frac{1}{T}\int_0^T {\left\{ {\mathop{{\int\!\!\!\!\!\int}\mkern-21mu \bigcirc} 
 {\left[ {f({x_{(t)}},{y_{(t)}}) \cdot \sum\limits_{i = 0}^p {\left( {C_p^i \cdot x_{(t)}^i{{\cos }^i}{\theta _{(t)}} \cdot y_{(t)}^{p - i}{{\sin }^{p - i}}{\theta _{(t)}}} \right)} } \right.} } \right.} \\
 &\cdot \left. {\left. {\sum\limits_{j = 0}^q {\left( {{{( - 1)}^j} \cdot C_q^j \cdot x_{(t)}^j{{\sin }^j}{\theta _{(t)}} \cdot y_{(t)}^{q - j}{{\cos }^{q - j}}{\theta _{(t)}}} \right)} } \right]d{x_{(t)}}d{y_{(t)}}} \right\}dt\\
&= \frac{1}{T}\int_0^T {\left\{ {\mathop{{\int\!\!\!\!\!\int}\mkern-21mu \bigcirc} 
 {\left[ {\sum\limits_{k = 0}^{p + q} {\left[ {\sum\limits_{i = 0}^k {\left( {C_p^i \cdot {{\cos }^i}{\theta _{(t)}} \cdot {{\sin }^{p - i}}{\theta _{(t)}} \cdot {{( - 1)}^{k - i}} \cdot C_q^{k - i} \cdot {{\sin }^{k - i}}{\theta _{(t)}} \cdot {{\cos }^{q + i - k}}{\theta _{(t)}}} \right)} } \right.} } \right.} } \right.} \\
&\left. {\left. {\left. { \cdot x_{(t)}^ky_{(t)}^{p + q - k}} \right] \cdot f({x_{(t)}},{y_{(t)}})} \right]d{x_{(t)}}d{y_{(t)}}} \right\}dt ~~~~~when ~~ i \le p~~and~~k - i \le q\\
 &= \sum\limits_{k = 0}^{p + q} {\left\{ {m_{k,p + q - k}^{(f)}\sum\limits_{i = 0}^k {\left[ {{{( - 1)}^{k - i}}C_p^iC_q^{k - i} \cdot \frac{1}{T}\int_0^T {\left( {{{\cos }^{q + 2i - k}}{\theta _{(t)}} \cdot {{\sin }^{p + k - 2i}}{\theta _{(t)}}} \right)} dt} \right]} } \right\}} 
\end{aligned}
\end{array}
\end{equation}
\end{justifying}

According the simplified result in the following definition (15), we can learn that geometric central moments of rotational-motion-blurred image and original image are linearly relative.
\begin{equation}
\begin{array}{c}
\begin{aligned}
m_{pq}^{(g)} &= \sum\limits_{k = 0}^{p + q} {\left\{ {m_{k,p + q - k}^{(f)} \cdot H(p,q,k)} \right\}} \\
H(p,q,k) &= \sum\limits_{i = 0}^k {\left[ {S1(p,q,k,i) \cdot S2(p,q,k,i)} \right]} \\
S1(p,q,k,i) &= \left\{ \begin{array}{l}
{( - 1)^{k - i}}C_p^iC_q^{k - i}~~~~~when~i \le p ~and~ k - i \le q\\
0~~~~~otherwise
\end{array} \right.\\
S2(p,q,k,i) &= \frac{1}{T}\int_0^T {\left( {{{\cos }^{q + 2i - k}}{\theta _{(t)}} \cdot {{\sin }^{p + k - 2i}}{\theta _{(t)}}} \right)} dt
\end{aligned}
\end{array}
\end{equation}

Depending on definition (15), we make further efforts to simplify derived results, so we have:
\begin{equation}
\begin{array}{c}
\begin{aligned}
S2(p,q,k,i) &= \frac{1}{T}\int_0^T {\left( {{{\cos }^{q + 2i - k}}{\theta _{(t)}} \cdot {{\sin }^{p + k - 2i}}{\theta _{(t)}}} \right)} dt\\
 &= \frac{1}{T}\int_0^T {\left( {{{\cos }^{p + q - x}}{\theta _{(t)}} \cdot {{\sin }^x}{\theta _{(t)}}} \right)} dt~~~(supposing~x = p + k - 2i)\\
 &= \frac{1}{{\omega T}}\int_0^{\omega T} {\left( {{{\cos }^{p + q - x}}\theta  \cdot {{\sin }^x}\theta } \right)} d\theta \\
 &= \frac{1}{{\omega T}}\int_0^{\omega T} {\left( {{{\cos }^{p + q}}\theta  \cdot {{\tan }^x}\theta } \right)} d\theta 
\end{aligned}
\end{array}
\end{equation}

Based on the definition, we can obtain different moments of any order. In this paper, we give central moments from 1st to 2th order.
\begin{equation}
\begin{array}{c}
\begin{aligned}
1st:m_{10}^{(g)} &= \frac{1}{{\omega T}}\left[ {m_{10}^{(f)} \cdot \sin (\omega T) + m_{01}^{(f)} \cdot \left( {1 - \cos (\omega T)} \right)} \right]\\
m_{01}^{(g)} &= \frac{1}{{\omega T}}\left[ {m_{10}^{(f)} \cdot \left( { - 1 + \cos (\omega T)} \right) + m_{01}^{(f)} \cdot \sin (\omega T))} \right]\\
2nd:m_{20}^{(g)} &= \frac{1}{{2\omega T}}\left[ {m_{20}^{(f)} \cdot \left( {\cos (\omega T)\sin (\omega T) + \omega T} \right) + 2m_{11}^{(f)} \cdot {{\left( {\sin (\omega T)} \right)}^2}} \right.\\
&\left. { + m_{02}^{(f)} \cdot \left( { - \cos (\omega T)\sin (\omega T) + \omega T} \right)} \right]\\
m_{11}^{(g)} &= \frac{1}{{2\omega T}}\left[ { - m_{20}^{(f)} \cdot {{\left( {\sin (\omega T)} \right)}^2} + 2m_{11}^{(f)} \cdot \cos (\omega T)\sin (\omega T) + m_{02}^{(f)} \cdot {{\left( {\sin (\omega T)} \right)}^2}} \right]\\
m_{02}^{(g)} &= \frac{1}{{2\omega T}}\left[ {m_{20}^{(f)} \cdot \left( { - \cos (\omega T)\sin (\omega T) + \omega T} \right) - 2m_{11}^{(f)} \cdot {{\left( {\sin (\omega T)} \right)}^2}} \right.\\
&\left. { + m_{02}^{(f)} \cdot \left( {\cos (\omega T)\sin (\omega T) + \omega T} \right)} \right]
\end{aligned}
\end{array}
\end{equation}

\subsection{Geometric moment invariants to rotational motion blur}
After screening out 6 Hu moment invariants and all rotation moment invaraints that are given in Mo’s paper\cite{ref10}, we find two absolute invaraints and ten relative invaraints that might have great invariace and robustness to rotational motion blur.
\begin{small}
\begin{equation}
\begin{array}{c}
\begin{aligned}
RMBMI - 1 &= m20 + m02\\
RMBMI - 2 &= m40 + 2m22 + m04\\
RMBMI - 3 &= \frac{{m01 \cdot m03 + m01 \cdot m21 + m10 \cdot m12 + m10 \cdot m30}}{{m_{10}^2 + m_{01}^2}}\\
RMBMI - 4 &= \frac{{m01 \cdot m12 + m01 \cdot m30 - m10 \cdot m03 - m10 \cdot m21}}{{m_{10}^2 + m_{01}^2}}\\
RMBMI - 5 &= \frac{{{{({m_{30}} + {m_{12}})}^2} + {{({m_{21}} + {m_{03}})}^2}}}{{m_{10}^2 + m_{01}^2}}\\
RMBMI - 6 &= \frac{{m02 \cdot m13 + m02 \cdot m31 - m04 \cdot m11 + m40 \cdot m11 - m13 \cdot m20 - m20 \cdot m31}}{{{{({m_{20}} - {m_{02}})}^2} + 4{m_{11}}^2}}\\
RMBMI - 7 &= \frac{{m{{01}^2} \cdot m13 + m{{01}^2} \cdot m22 - m01 \cdot m04 \cdot m10 + m01 \cdot m10 \cdot m40 - m{{10}^2} \cdot m13 - m{{10}^2} \cdot m31}}{{m{{01}^2} \cdot m11 - m01 \cdot m02 \cdot m10 + m01 \cdot m10 \cdot m20 - m{{10}^2} \cdot m11}}
\end{aligned}
\end{array}
\end{equation}
\end{small}

\clearpage
\section{Conclusions}
In this paper, we first combine the concept of geometric moment and motion blur degradation model; then we prove that geometric moments of blurred image and original image are linearly related. Depending on this property, we give an approach of analysing whether an existing moment-based feature is invariant to motion blur. Finally, we find some geometric moment invariants are invariants to not only spatial transform but also motion blur. Meanwhile, we test invariance and robustness of these invariants using synthetic and real blur image datasets.

\section*{Acknowledgements}
This work was partly funded by National Key R\&D Program of China (no.2017YFB1002703), National Key Basic Research Planning Project of China (no. 2015CB554507) and National Natural Science Foundation of China (no.61227802 and 61379082).

\end{document}